\documentclass[12pt]{article}
\usepackage{latexsym}
\usepackage{graphicx,subfigure}
\usepackage{amsmath,amssymb,latexsym,amsfonts}
\usepackage{cite}
\usepackage{hyperref}
\usepackage[mathscr]{eucal}

\newtheorem{thm}{Theorem}
\newtheorem{lem}[thm]{Lemma}

\newtheorem{rmk}{Remark}
\newtheorem{dfn}{Definition}
\newtheorem{crl}[thm]{Corollary}

\begin{document}

\begin{center}
\textbf{\Large On the spectral stability of soliton-like solutions to a non-local hydrodynamic-type model  \footnote{The paper is published in Communications in Nonlinear Science and Numerical Simulation (2019), DOI:10.1016/j.cnsns.2019.104998}} \vspace{0.5 cm}

Vsevolod A.~Vladimirov$^\ddagger$ \footnote{e-mail:
\url{vsevolod.vladimirov@gmail.com}},
Sergii Skurativskyi$^\dagger$ 
\footnote{e-mail:
\url{skurserg@gmail.com}},

 \vspace{0.5 cm}

$^\ddagger$Faculty of Applied Mathematics,
AGH University of Science and Technology,
Mickiewicz Avenue 30, 30059 Krak\'{o}w, Poland

$^\dagger$Subbotin Institute of Geophysics, NAS of Ukraine, Kyiv, Ukraine

\end{center}

\begin{quote} \textbf{Abstract. }{\small A model of nonlinear elastic medium with internal structure is considered. The medium is assumed to contain cavities, microcracks or blotches of substances that differ sharply in physical properties from the base material. To describe the wave processes in such a medium, the averaged values of physical fields are used. This leads to nonlinear evolutionary PDEs, differing from the classical balance equations. The system under consideration possesses a family of invariant soliton-like solutions. These solutions are shown to be spectrally stable under certain restrictions on the parameters.
}
\end{quote}

\begin{quote} \textbf{Keyword: }{\small 
 Nonlocal hydrodynamic-type model; Hamiltonian formulation; Multisymplectic formulation; Spectral stability of soliton-like solutions
}
\end{quote}

\vspace{0.5 cm}

\section{Introduction}

Basing on \cite{Perlings}, it was proposed in \cite{VMSS} the following model describing a nonlinear elastic medium with internal inclusions, cavities or microcracks:
\begin{eqnarray}
u_t+\frac{1}{\nu+2}\partial_x\,\left(\beta+\sigma\partial^2_x\right) \rho^{\nu+2}   =0,  \label{Perl_1A}\\
\rho_t+\rho^2\,u_x=0, \label{Perl_1B}
\end{eqnarray}
where $\beta>0,$ $\sigma\neq 0$, $\nu>-1$.
In paper \cite{VMSS} a family of traveling wave (TW) solutions satisfying the system (\ref{Perl_1A})-(\ref{Perl_1B}) is investigated and conditions are formulated under which the soliton-like TW solutions exist. Depending on the sign of the parameter $\sigma$, the soliton-like solutions describe the waves of compression (when $\sigma>0$) or the waves of rarefaction (corresponding to $\sigma<0$). Additionally, in paper \cite{VMSS} the stability of soliton-like solutions is investigated, based on the numerical study of the Evans function \cite{Evans3,Evans4,KaPromis}. Unfortunately, it is impossible to trace the analytical relationship between stability and the parameters' values, using numerical studies. However, the possibility to get the analytical results appears in the case when the model under investigation allows a Hamiltonian description. The rigorous studies of stability properties of TW solutions to various nonlinear models have been carried out in papers \cite{Benjamin,PW,BriDerks_97,BriDerks_02,AS,Zumburn_2006,pap12} in which some general results are formulated concerning the properties of spectral operators, that make it possible to estimate a number of unstable modes. In some cases, it is possible to completely eliminate the presence of unstable modes by investigating the function of the spectral parameter put forward by Evans \cite{Evans3,Evans4} and bearing his name. In the general case, this function can only be calculated numerically, but for our purposes it is sufficient to study its asymptotic properties, as well as its behavior at the origin, which can be done analytically. Following this way, we succeeded in obtaining restrictions on the parameters which give the sufficient conditions for the spectral stability of soliton-like TW solutions. The structure of this work is as follows. In Section 2 we pass from the system  (\ref{Perl_1A})-(\ref{Perl_1B}) to the equivalent system, having nice a Hamiltonian representation, and state the conditions assuring the existence of soliton-like TW solutions. In Section 3 we concentrate on the analysis of the spectral stability. We study the linearized system obtained by varying the soliton-like solution. Using the approach based on the Sturm-like theorems, we first estimate the maximal number of unstable modes and then formulate the conditions, which guarantee their absence. In this section, we use the technique based on a somewhat cumbersome multisymplectic representation of the Hamiltonian system. So in order not to clutter the main text, we provide some technical details in Appendices A and B. Finally, in Section 4 we summarize the results obtained and discuss further research.

\section{Hamiltonian representation and soliton-like solutions}

Let us consider the following substitution
\begin{eqnarray}\label{nloctransf}
u=\left(\gamma-\kappa\partial^2_x\right) w, \qquad   \eta=\frac{1}{\rho},
\end{eqnarray}
where $\gamma=\beta/(\nu+2)>0,$ $\kappa=-\sigma/(\nu+2)>0.$
Inserting (\ref{nloctransf}) into  (\ref{Perl_1A})-(\ref{Perl_1B}), we get the equations
\begin{eqnarray*}
-\frac{1}{\eta^2} \left\{\eta_t-\left(\gamma-\kappa\partial^2_x\right) w_x   \right\}   =0, \label{UP1} \\
 \left(\gamma-\kappa\partial^2_x\right) \left\{w_t+\partial_x\,\eta^{-(\nu+2)} \right\} =0.  \label{UP2}
\end{eqnarray*}
Under the above assumptions the operator 
$
\mathscr{P}=\gamma-\kappa\partial^2_x
$ 
is invertible, so we can rewrite the initial system in the following equivalent form
\begin{eqnarray}
w_t=-\partial_x\,\eta^{-(\nu+2)},  \label{MGM1} \\
 \eta_t=\left(\gamma-\kappa\partial^2_x\right) w_x. \label{MGM2} 
\end{eqnarray}
By direct verification, one can get convinced that the system (\ref{MGM1})-(\ref{MGM2}) admits the Hamiltonian representation
\begin{equation}\label{Perl_Hamilt_1}
U_t=\partial_x \,\left(\begin{array}{cc} 0 & 1 \\ 1 & 0 \end{array} \right) \delta H=J\cdot \delta H,
\end{equation}
where $U=\left(w,\,\,\eta   \right)^{tr},$
\[
H= \int_{-\infty}^{+\infty}\left\{\frac{1}{2}\left[\gamma\,w^2+\kappa w_x^2\right]- \int_{\eta_{\infty}}^\eta\,\left[p(\xi)-p(\eta_{\infty})  \right]d\,\xi\right\}\,d\,x,
\]
$0<\eta_{\infty}=\lim\limits_{|x| \to \infty  }\eta(t,x),\,$  $p(\xi)=1/\xi^{\nu+2}.$ 

In the sequel we will be interested in a family of the TW solutions $w=w_s(z),$ $\eta=\eta_s(z),$ where $z=x-s\,t,$ so we rewrite the system (\ref{Perl_Hamilt_1}) with the traveling wave coordinates $\bar t=t,\,\,\,\bar z=x-s\,t$:
\begin{equation}\label{Hamilt_TW}
U_{\bar t}=\partial_{\bar z} \,\left(\begin{array}{cc} 0 & 1 \\ 1 & 0 \end{array} \right) \delta ( H+s\,Q),
\end{equation}
where
\[
Q=\int_{-\infty}^{+\infty} w (\eta-\eta_{\infty})\,d\,z
\]
is the generalized momentum (we omit bars over the independent variables in what follows). Since in the new coordinates the TW solutions are stationary, they satisfy the system
\begin{equation}\label{Var_TW}
\partial_{z} \,\left(\begin{array}{cc} 0 & 1 \\ 1 & 0 \end{array} \right)\,\delta ( H+s\,Q)|_{w_s (z),\, \eta_s (z)}=0.
\end{equation}

Now we are going to formulate the conditions which guarantee the existence of homoclinic solutions representing the solitary waves. The system (\ref{Var_TW}) can be rewritten  as follows:
\begin{equation}\label{VarTW1}
\partial_z\left\{s  \,w_s-\eta_s^{-(\nu+2)}+\eta_\infty^{-(\nu+2)}   \right\}=0,
\end{equation}
\begin{equation}\label{VarTW2}
\partial_z\left\{\left(\gamma-\kappa\,\partial_{z}^2   \right) w_s+s\left(\eta_s-\eta_\infty   \right)   \right\}=0.
\end{equation}
Integrating these equations from $-\infty$ to $z$ and taking into account the asymptotics
\begin{equation}\label{asympt}
\lim\limits_{|z| \to \infty  } \eta_s(z)=\eta_\infty,  \quad \lim\limits_{|z| \to \infty  } w_s(z)=0,
\end{equation}
we get the system
 \begin{equation}\label{varsol1}
s  \,w_s+\eta_{\infty}^{-(\nu+2)} - \eta_s^{-(\nu+2)}   =0,
\end{equation}
\begin{equation}\label{varsol2}
\left(\gamma-\kappa\,\partial_{z}^2   \right) w_s+s\left(\eta_s-\eta_\infty   \right)=0.
\end{equation}
Using Eq.~(\ref{varsol1}), we can eliminate the function $w_s$ from Eq.~(\ref{varsol2}). Next, introducing a new variable $\theta=\eta^\prime_s$ and using the integrating factor $\varphi=\eta_s^{-(\nu+3)}$, we can rewrite Eq. (\ref{varsol2}) in the form of a Hamiltonian system
\begin{equation}\label{HDS}\left\{ \begin{array}{l}
\frac{d}{d\,T}\eta_s=\theta\kappa\,(\nu+2)\,\varphi^2=\mathcal{H}_\theta, \\
\frac{d}{d\,T} \theta=\varphi\left\{\kappa(\nu+2)(\nu+3) \theta^2\,\eta_s^{-(\nu+4)}- \right.\\ 
\left.   \hspace{25mm}-\left[s^2\left(\eta_s-\eta_\infty   \right)+\gamma\left(\frac{1}{\eta_s^{\nu+2}}-\frac{1}{\eta_\infty^{\nu+2}}   \right)\right]   \right\}= -\mathcal{H}_{\eta_s},  \end{array} \right.   
\end{equation} 
where $\frac{d}{d\,T}=\kappa\,(\nu+2)\,\varphi^2\,\frac{d}{d\,z},$
$
\mathcal{H}=E_k(\eta_s,\,\theta)+V(\eta_s),
$
\[
E_k(\eta_s,\,\theta)=\frac{\kappa}{2}\,(\nu+2)\,\eta^{-2(\nu+3)}\theta^2
\]
is the kinetic energy, while 
\[
V(\eta_s)=s^2\left[ \frac{\eta_\infty}{(\nu+2)\eta_s^{\nu+2}}-\frac{1}{(\nu+1)\eta_s^{\nu+1}}\right]+\gamma \left[\frac{1}{(\nu+2)(\eta_s\eta_\infty)^{\nu+2}}-\frac{1}{2\,(\nu+2)\eta_s^{2(\nu+2)}}   \right]
\]
is the potential energy.

Using the well-known properties of two-dimensional Hamiltonian systems \cite{AndrKhajk}, we can perform exhaustible qualitative analysis of the system (\ref{HDS}). It is evident, that all stationary points of this system are placed on the horizontal axis. The coordinate $\eta$ of a stationary points satisfies the equation
\begin{equation}\label{etastac}
s^2 \left(\eta_\infty-\eta\right)=\gamma\left(\frac{1}{\eta^{\nu+2}}-\frac{1}{\eta_\infty^{\nu+2}}   \right).
\end{equation}
Eq. (\ref{etastac}) is fulfilled when $\eta=\eta_\infty,$ so $(\eta_\infty,\,0)$ is the stationary point. We are looking for soliton-like solutions satisfying the asymptotic conditions (\ref{asympt}) and thus corresponding to the phase trajectories bi-asymptotic to the stationary point $(\eta_\infty,\,0)$ which must be a saddle. This is so if the eigenvalues of the Jacobi matrix
\begin{equation}\label{Jacmatr}
\mathscr{R}|_{\eta=\eta_\infty,\,\theta=0}=\left( \begin{array}{cc} 0 & \varphi(\eta_\infty)\,\kappa (\nu+2)\,\eta_\infty^{-(\nu+3)} \\  -\varphi(\eta_\infty) \left[s^2-\gamma\,(\nu+2)\,\eta_\infty^{-(\nu+3)}\right] & 0 \end{array} \right),
\end{equation} 
are real numbers of different signs, or, in other words, if the following inequality holds:
\begin{equation}\label{rightineq}
s^2<\gamma\,(\nu+2)\,\eta_\infty^{-(\nu+3)}.
\end{equation}

\begin{figure}[h]
\centering
 \includegraphics[totalheight=2.5 in]{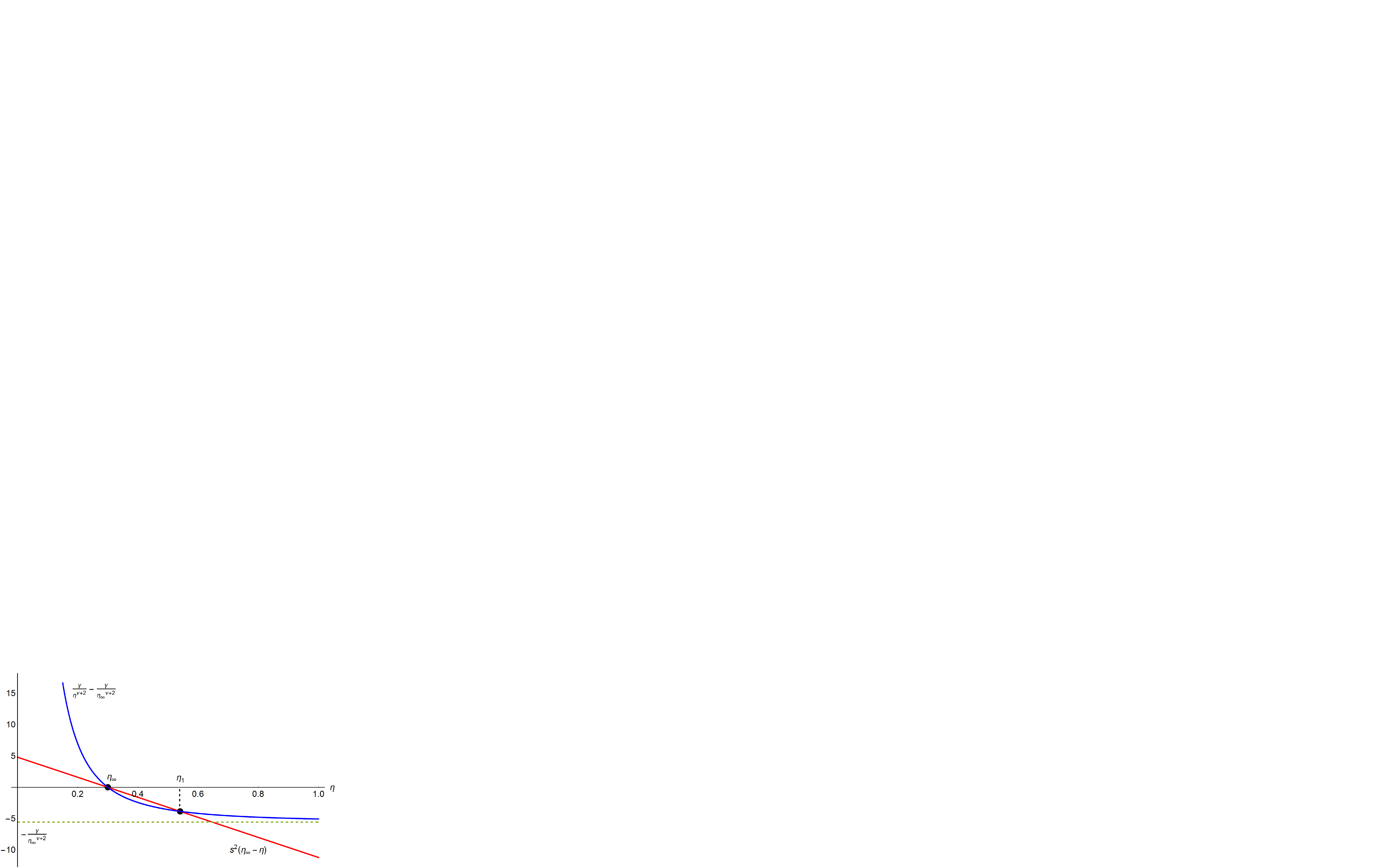}
 \caption{Graphical solution of Eq.~(\ref{etastac}).}\label{Fig:1}
\end{figure}

The fulfillment of the condition (\ref{rightineq}) also implies the existence of a second solution of Eq.~(\ref{etastac}), located to the right of  $\eta_\infty$, see Fig. \ref{Fig:1}. The second solution, which we denote by $\eta_1$, satisfies the inequality
\[
s^2>\gamma\,(\nu+2)\,\eta_1^{-(\nu+3)}.
\]
Under the above condition, the eigenvalues of the Jacobi matrix $\mathscr{R}$  in the stationary point $(\eta_1,\,0)$ are pure imaginary, so it is a center.

\begin{figure}[h]
  \centering
  \subfigure{\includegraphics[totalheight=2 in]{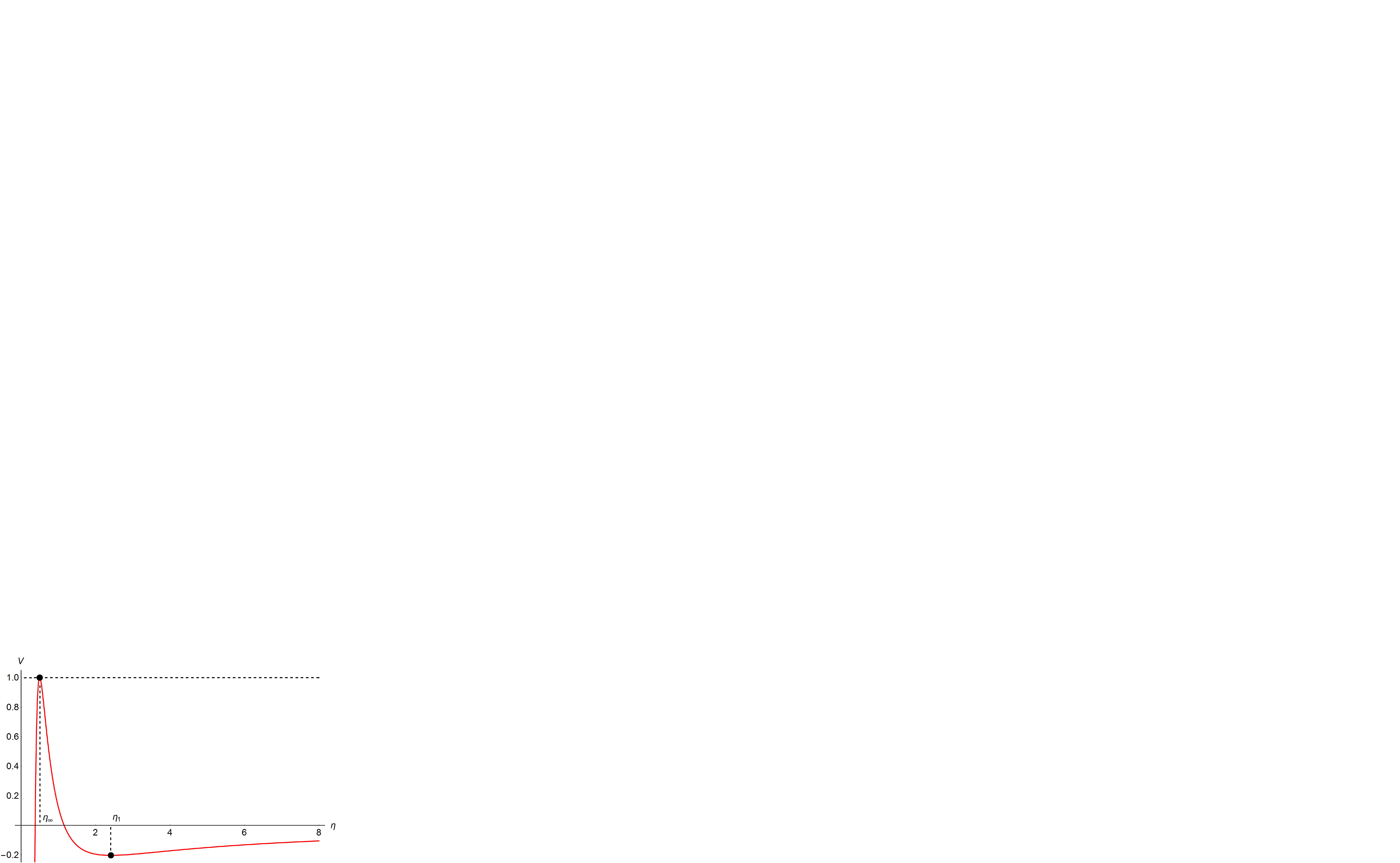}}\quad 
	\subfigure{\includegraphics[totalheight=2 in]{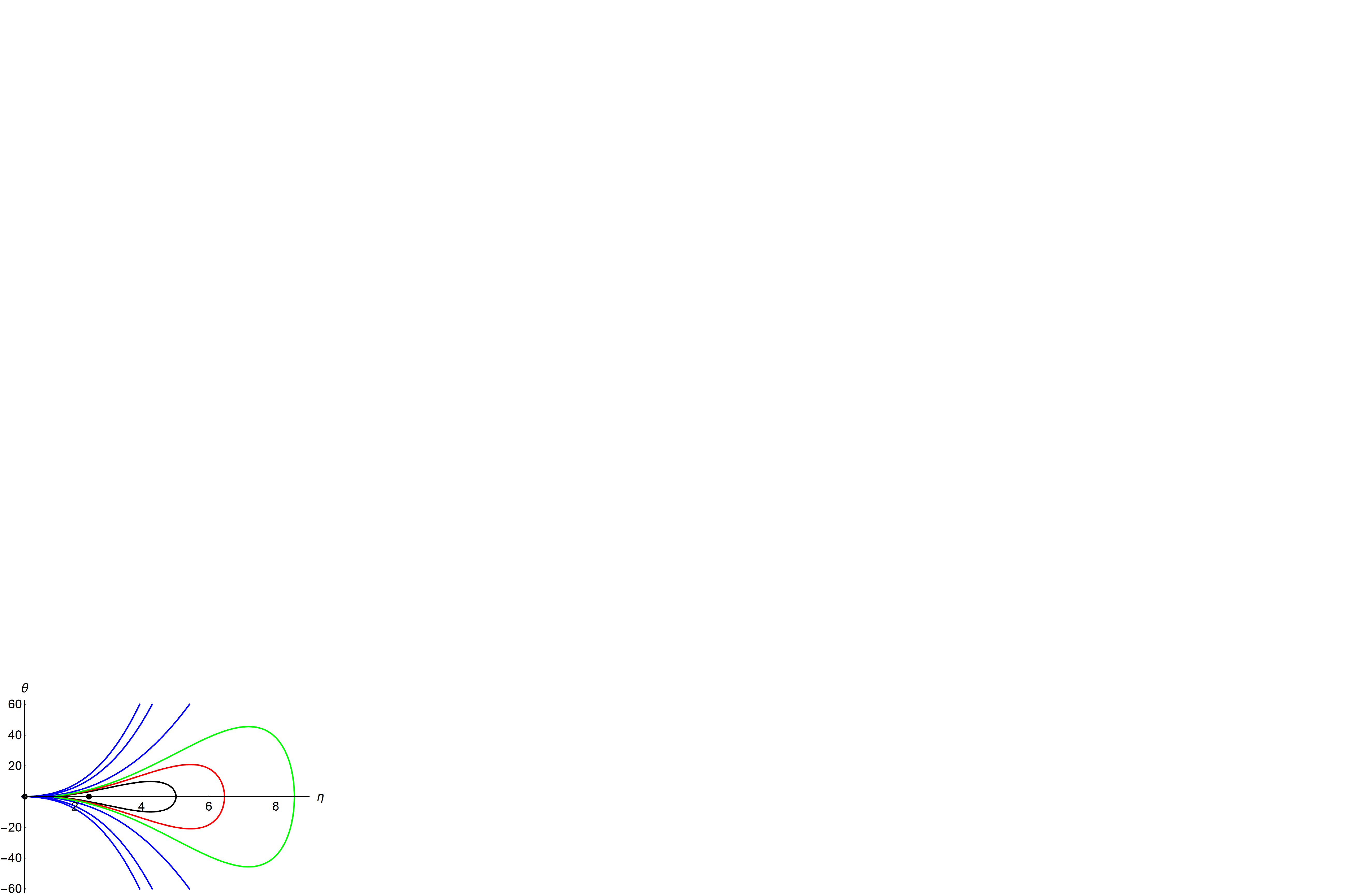}}
  \caption{Graph of potential energy $V(\eta)$, case $V(\eta_\infty)>0$ (left panel) and the corresponding phase portrait (right panel). All the trajectories shown represent the periodic solutions. }\label{Fig:2}
\end{figure}

\begin{figure}[h]
 \centering
  \subfigure{\includegraphics[totalheight=2 in]{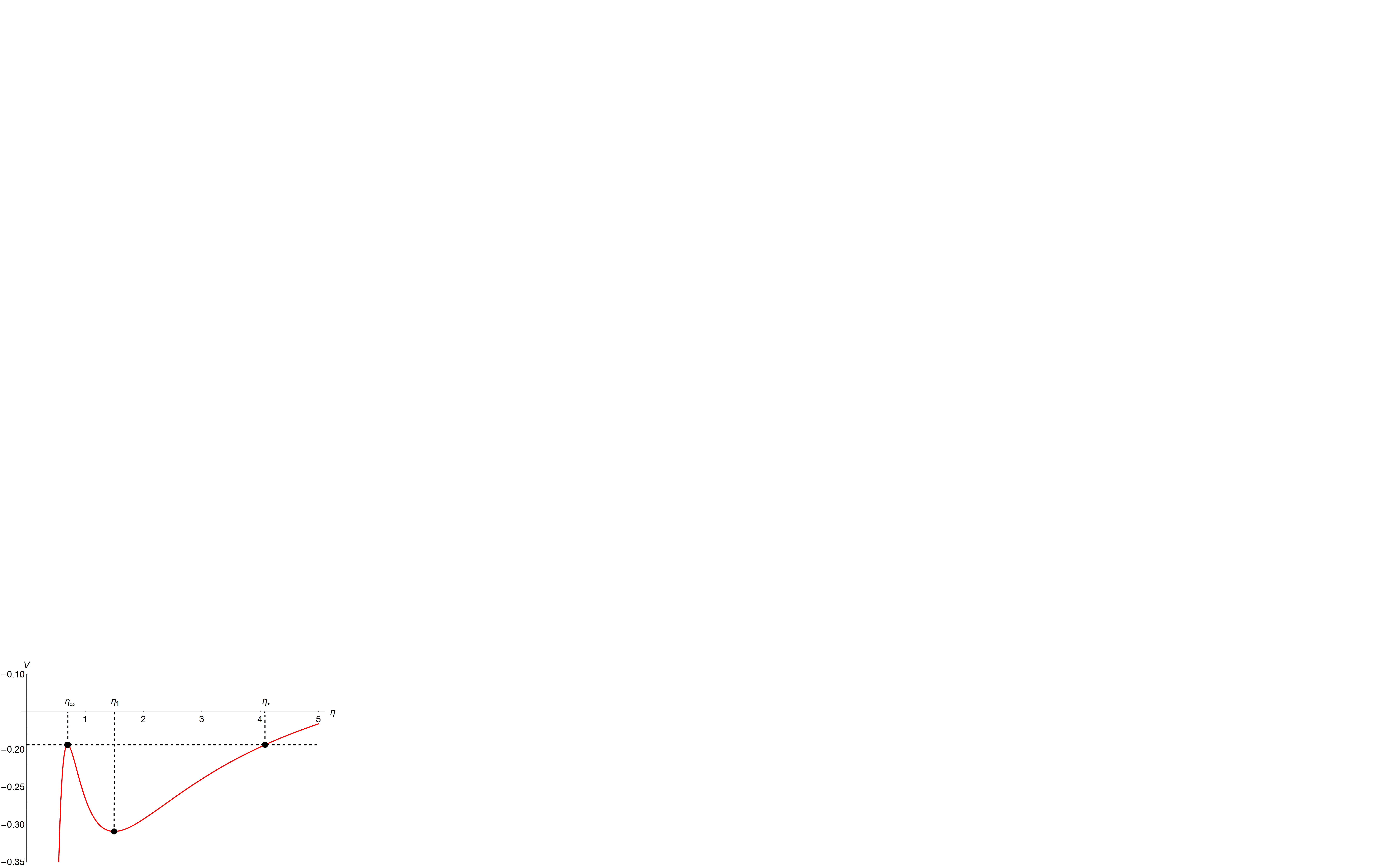}}\quad
	\subfigure{\includegraphics[totalheight=2 in]{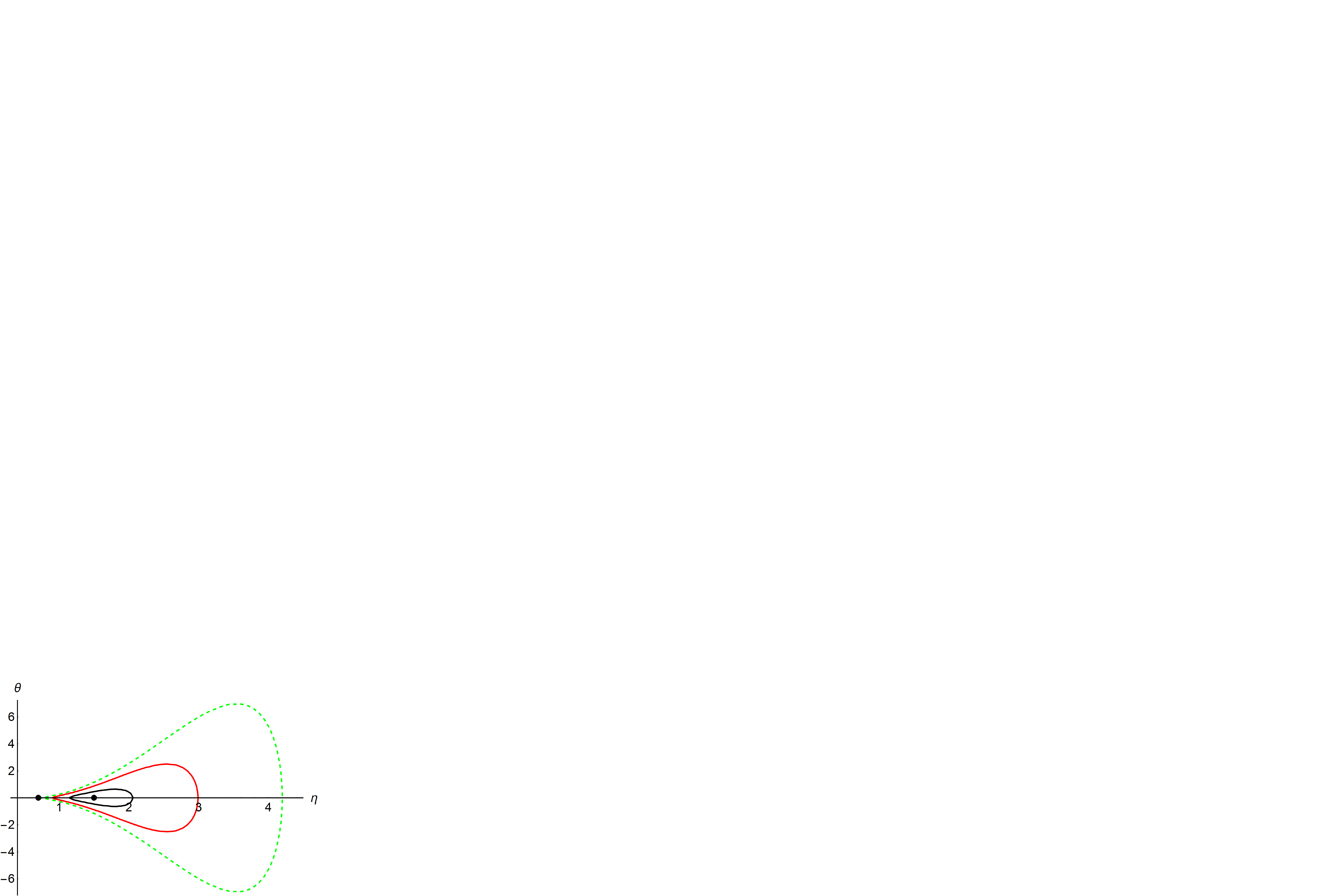}}
  \caption{Graph of potential energy $V(\eta)$, case $V(\eta_\infty)<0$ (left panel) and the corresponding phase portrait (right panel). Dashed line corresponds to the homoclinic loop; solid lines represent the periodic solutions. }\label{Fig:3}
\end{figure}

An extra condition, which, together with (\ref{rightineq}), guarantees the existence of the homoclinic loop is connected with the general features of two-dimensional Hamiltonian systems. As is well-known \cite{AndrKhajk}, the Hamiltonian function remains constant on the phase trajectories of the Hamiltonian system. The potential energy of the system (\ref{HDS}) has exactly two local extrema, namely, the local maximum at the point $\eta_\infty$ and the local minimum at the point $\eta_1.$ It is easy to check that $\lim\limits_{\eta \to +\infty  } V(\eta)=0$ and, depending on the values of the parameters, two distinct configurations occur. If $V(\eta_\infty)\,\geq\,0,$ then the level line passing through the point of local maximum  unlimitedly extends  to the  right  without intersecting the graph  of the function $V(\eta)$ (see Fig. \ref{Fig:2}, left panel). In this case the stable and unstable saddle separatrices do not form a closed loop, and the region of the phase plane $(\eta,\,\theta)$ bounded by these separatrices   is filled with the periodic trajectories spreading up to infinity. If $V(\eta_\infty)\,<\,0,$ then the level line passing through the point of local maximum intersects the graph of the function $V(\eta)$ at a point $\eta_{*}$, $\eta_1<\eta_{*}<\infty$ (see Fig. \ref{Fig:3}).

\noindent
The phase trajectory cannot have a coordinate $\eta$ greater than $\eta_*$, so at the point $(\eta_*,\,0)$  the outgoing trajectory of the saddle point $(\eta_\infty,\,0)$ is reflected from the horizontal axis. Since the Hamiltonian function is not changed under the replacement $\theta$ by $-\theta$, the straight and reflected trajectories are symmetric with respect to the horizontal axis and form a single homoclinic trajectory bi-asymptotic to the saddle.  Condition $V(\eta_\infty)\,<\,0,$ together  with the condition which guarantees that the stationary point $(\eta_\infty,\,0)$ is a saddle, forms the pair of inequalities
\begin{equation}\label{hclexist}
\frac{\beta (\nu+1)}{2(\nu+2) \eta_\infty^{\nu+3}}<s^2<\frac{\beta}{\eta_\infty^{(\nu+3)}}
\end{equation} 
assuring the presence of soliton-like regimes  in the set of TW solutions.

\begin{rmk}\label{existHCL}
If we introduce the parameter $\eta_{0,\,\infty}=s^{2/(\nu+3)}\,\eta_\infty$, then the velocity $s$ will be eliminated from (\ref{hclexist})  which acquires the following form
\begin{equation}\label{hclexist0}
\frac{\beta (\nu+1)}{2(\nu+2) }<\eta_{0,\,\infty}^{\nu+3}<{\beta}.
 \end{equation}
In what follows we treat the parameter $\eta_{0,\,\infty}$ as independent of $s.$
\end{rmk}

To conclude, let us note that the conditions presented in (\ref{hclexist}) coincide with those obtained in paper \cite{VMSS} after the performance of substitution $R_1=\eta_\infty^{-1}.$

\section{Spectral stability of the soliton-like solutions}

\subsection{Restrictions on the number of unstable modes}

In order to study the stability of solitary wave solutions, the following set of perturbations is considered:
\begin{equation}\label{perspec}
\left(\begin{array}{c} w(t,\,z) \\ \eta(t,\,z) \end{array} \right)=
\left(\begin{array}{c} w_s(z) \\ \eta_s(z) \end{array} \right)+
\varepsilon\,e^{\lambda\,t} \left(\begin{array}{c} M(z) \\ N(z) \end{array} \right).
\end{equation}

Inserting (\ref{perspec}) into (\ref{Hamilt_TW}), we get, up to $O(\varepsilon^2)$ the eigenvalue problem
\begin{equation}\label{eigenvalprobl}
\lambda\,\left(\begin{array}{c} M(z) \\ N(z) \end{array} \right)=J \, \mathcal{L}^s\,\left(\begin{array}{c} M(z) \\ N(z) \end{array} \right):=\mathbb{L}\,\left(\begin{array}{c} M(z) \\ N(z) \end{array} \right),
\end{equation}
where
\begin{equation}\label{Ls}
\mathcal{L}^s=\delta^2\,\left( H+s\,Q  \right)|_{w_s,\,\eta_s}=
\left(\begin{array}{cc} \gamma-\kappa\,\partial_z^2 &  s \\ s  & (\nu+2)/{\eta^{\nu+3}_s(z)} \end{array} \right).
\end{equation}
We denote the spectrum of the operator $\mathbb{L}$ by $\sigma(\mathbb{L})$ and accept the following definition: 

\begin{dfn}
The soliton-like solution $U_s(z)=\left(w_s(z),\,\eta_s(z)\right)^{tr}$ is said to be spectrally stable if the intersection of $\sigma(\mathbb{L})$ with the positive half-plane $\mathbb{C}^+$ of the complex plane is empty.
\end{dfn}

In this section  the following statement will be proved:

\begin{thm}
The set $\sigma(\mathbb{L})\cap\,\mathbb{C}^+$ consists of at most one isolated point $\lambda_0.$ If  $\sigma(\mathbb{L})\cap\,\mathbb{C}^+$ is nonempty, then $\lambda_0$ is a real positive number. 
\end{thm}

The proof of this theorem  is based on a number of auxiliary statements. Some of them are sufficiently general and applicable to a wide class of spectral problems. To begin with, let us localize the essential spectrum $\sigma_{ess}(\mathbb{L}).$ In the case under consideration it coincides with the spectrum of the limiting operator \cite{D_Henry,KaPromis}
\[
\mathbb{L}_\infty=\mathbb{L}_{\pm\infty}=\lim\limits_{|z| \to \infty  }J \cdot \mathcal{L}^s=
J\cdot \left(\begin{array}{cc} \gamma-\kappa\,\partial_z^2 &  s \\ s  & \frac{\nu+2}{\eta_\infty^{\nu+3}} \end{array} \right).
\]
The spectrum of the operator $\mathbb{L}_\infty,$ having the constant coefficients, coincides with the set
\[
\sigma_{ess}(\mathbb{L})=\left\{\lambda\in \mathbb{C}: \det\left(\begin{array}{cc} - i\,\xi\,s-\lambda &  - i\,\xi(\nu+2)/\eta_\infty^{\nu+3} \\ - i\,\xi (\gamma+\kappa\xi^2)  & - i\,\xi\,s-\lambda \end{array} \right)=0, \,\,\xi\in\,\mathbb{R}      \right\}.
\]    
The set of possible values of the spectral parameter $\lambda$ is given by the formula
\[
\lambda=- i \xi s \pm i \sqrt{\xi^2 (\nu+2)\,(\gamma+\kappa \xi^2)/\eta_\infty^{\nu+3}}, \quad \xi\in\,\mathbb{R}.
\] 
It coincides with the imaginary axis. 
Next, the following general statement is applied to our problem  (cf with \cite{KaPromis}).

\begin{lem}
The point spectrum  $\sigma_{pt}(\mathbb{L})$ is symmetric with respect to the coordinate axes, that is, if $\lambda \in \sigma_{pt}(\mathbb{L}), $ then simultaneously $-\lambda,$ and $\pm\lambda^{*}$ belong to the point spectrum of the operator $\mathbb{L}.$
\end{lem}
{\bf Proof.}
Suppose that $\lambda\in\sigma_{pt}(\mathbb{L}),$ $\psi\,\in L^2(\mathbb{R})$ is the eigenvector corresponding to $\lambda$. Then
\[
\left(\mathbb{L}\,\psi\right)^{*}=\mathbb{L} \psi^{*}=\lambda^{*} \psi^{*},
\]
hence $\lambda \in \sigma_{pt}(\mathbb{L})$ implies $\lambda^{*} \in \sigma_{pt}(\mathbb{L})$.
Next,
if $\mathbb{L}\,\psi=J\cdot \mathcal{L}^s\,\psi=\lambda\,\psi,$ then
\[
\left(\psi | J\cdot \mathcal{L}^s \psi \right)=
\left(\psi | \lambda\, \psi \right)=
\lambda\,\left(\psi |  \psi \right)=
\left(\lambda^{*}\,\psi |  \psi \right).
\] 
On the other hand, 
\[
\left(\psi | J\cdot \mathcal{L}^s \psi \right)=-\left(J\,\psi | \mathcal{L}^s \psi \right)=
-\left( \mathcal{L}^s \cdot J\,\psi | \psi \right),
\]
hence 
$
\mathcal{L}^s \cdot J\,\psi=-\lambda^{*}\,\psi, 
$
which implies the equality $\mathbb{L} \left(J \,\psi\right)=- \lambda^{*}\left(J\,\psi\right).$ Thus, $\lambda \in \sigma_{pt}(\mathbb{L})$ implies $-\lambda^{*} \in \sigma_{pt}(\mathbb{L})$ and similarly  $\lambda^{*} \in \sigma_{pt}(\mathbb{L})$ implies $-\lambda \in \sigma_{pt}(\mathbb{L}).$ 

The next statement, borrowed from the paper \cite{PW} is the following.

\begin{thm}
Suppose that $J$ is a skew-symmetric operator while $\mathcal{L}^s$ is self-adjoint. Suppose in addition that  $\mathcal{L}^s$ has exactly $k$ strictly negative eigenvalues, counting multiplicities and $k<\infty.$ Then $\mathbb{L}=J\cdot\,\mathcal{L}^s$ has at most $k$ eigenvalues in the right half-plane of the complex plane.
\end{thm}

So, all the auxiliary assertions needed have been formulated, and we can now concentrate on estimating the number of discrete eigenvalues of the operator  $\mathcal{L}^s$ lying on the negative  semiaxis $\mathbb{R}^-$. Thus, we consider the spectral problem $\mathcal{L}^s\,\left(M,\,N\right)^{tr}=\mu\,\left(M,\,N\right)^{tr},$ which can be presented as follows:
\begin{equation}\label{spectrLs}
\left\{ \begin{array}{l} (\gamma-\kappa \partial_z^2)\,M+s\,N=\mu\,M, \\ \\
s\,M+\frac{\nu+2}{\eta^{\nu+3}_s(z)}\,N=\mu\,N. \end{array} \right.
\end{equation}
Note that if we put in  (\ref{spectrLs}) $\mu=0$ and make the replacement $M=w'_s$, $N=\eta'_s$ then as a result we get the system  (\ref{varsol2})-(\ref{varsol1}). Hence the following statement is true:

\begin{lem}\label{eigenzero}
$U_s^\prime=(w_s^\prime,\,\eta_s^\prime)^{tr}$ is the eigenvector of the operator $\mathcal{L}^s,$ corresponding to the eigenvalue $\mu=0.$
\end{lem}

Now, using the second equation of the system (\ref{spectrLs}), we can express the function $N$ as follows:
\begin{equation}\label{ansatzN}
N=s\,M\,\left(\mu-\frac{\nu+2}{\eta^{\nu+3}_s(z)}   \right)^{-1}.
\end{equation}
Inserting (\ref{ansatzN}) into the first equation of the system (\ref{spectrLs}), we get the following generalized eigenvalue problem:
\begin{equation}\label{geneigen}
\kappa \frac{d^2}{d\,z^2} \,M=\left[\gamma-\mu+\frac{s^2}{\mu-\frac{\nu+2}{\eta^{\nu+3}_s(z)}}   \right] \,M.
\end{equation}
From lemma \ref{eigenzero} we immediately obtain the following:

\begin{crl}
Function $M(z)=w^\prime_s(z)$ is the eigenvector of the generalized spectral problem (\ref{geneigen}) corresponding to the eigenvalue $\mu=0.$
\end{crl}

Now, let us consider the Wronskian
\begin{equation}\label{WR}
W(z)=M_1^\prime(z)\,M_2(z)-M_2^\prime(z)\,M_1(z),
\end{equation}
on a set $(a,\,b)\in \mathbb{R}$  (finite or infinite), where $\left\{M_i\right\}_{i=1}^2$ are solutions of Eq.~(\ref{geneigen}) corresponding to the eigenvalues $\mu_i.$  Taking the derivative of (\ref{WR}) with respect to $z$ and then integrating the expression obtained we get

\[W(\xi)\,|_{\xi=a}^{\xi=z}=\int_a^z W^\prime(\xi)\,d\,\xi,\]
which after some manipulation attains the form
\begin{equation}\label{intWR}
W(z)-W(a)=\frac{\mu_2-\mu_1}{\kappa}\int_a^z M_1(\xi)\,M_2(\xi) \Phi(\xi)\,d\,\xi,
\end{equation}
where
\[
\Phi(\xi)=1+
\frac{s^2\,\eta_s^{2(\nu+3)}(\xi)}{\left[(\nu+2)-\mu_1\,\eta_s^{\nu+3}(\xi)  \right]\,\left[(\nu+2)-\mu_2\,\eta_s^{\nu+3}(\xi)  \right]}.
\]
Let us note, that $\Phi(\xi)>0$ when $\mu_i\,\leq\,0,\,\,i=1,\,2.$

\begin{lem}\label{Sturm_1}
Let us assume that $\mu_1<\mu_2\leq 0$ are the eigenvalues while $M_1(z),\,\,M_2(z)$ are the  corresponding eigenfunctions of the generalized spectral problem (\ref{geneigen}), $c\,\in\,(a,\,b)$ and the following conditions hold:
\begin{itemize}
\item
$\lim\limits_{z \to a+0  } M_1(z)=\lim\limits_{z \to a+0  } M_2(z)=0$;
\item
$M_2|_{(a,c)}>0; \quad \exists\, \epsilon>0: \,\,M_1|_{(a,a+\epsilon)}>0.$
\end{itemize} 
Then  $M_1\,|_{(a,c)}>0$. If in addition the conditions
\begin{itemize}
\item
$M_2(c)=0, \quad M_2^\prime(c)<0$
\end{itemize} 
are fulfilled, then $M_1(c)>0.$
\end{lem}

{\bf Proof.} 
The proof of the first statement: let there exists $d\in (a,\,c)$ such that $M_1(d)=0$ and $M_1^\prime(d)<0$ (we'll assume that $d$ is the first point at which $M_1(z)$ intersects the horizontal axis). Then the function (\ref{intWR}) is growing and non-negative on $(a,\,d).$ On the other hand, it appears from (\ref{WR}), that under the above assumption $W(d)=M_1^\prime(d)\,M_2(d)<0.$ The resulting contradiction eliminates this possibility. Now let us address the second statement. It appears from (\ref{intWR}) that $W(c)>0.$ Using the additional assumptions, we conclude from (\ref{WR}) that $W(c)=-M_2^\prime(c)\,M_1(c).$ But this expression can be positive only if  $M_1(c)>0.$  

\begin{lem}\label{Sturm_2}
We use the same assumptions as in the first part of the lemma \ref{Sturm_1}. In addition, we assume that
\begin{itemize}
\item
$M_2(c)=0; \quad M_2^\prime(c)<0; \quad \exists\,\, e>c\,: M_2|_{(c,\,e)}<0;$
\item $\lim\limits_{z \to e-0  } M_1(z)=\lim\limits_{z \to e-0  } M_2(z)=0.$
\end{itemize}
Then $M_1\,|_{(c,e)}>0.$
\end{lem}

{\bf Proof.}
The lemma is proved by contradiction. Assume that $M_1(z)$ intersects the horizontal axis $OZ$ for the first time at some point $f\in (c,\,e).$ Then $M_1(f)=0, \quad M_1^\prime(f)<0,$  hence $W(f)=M_1^\prime(f) M_2(f)-M_2^\prime(f) M_1(f)>0.$ Let us also make an additional assumption that $M_1(z)$ does not have intersections with the horizontal axis on the segment $(f,\,e).$ Then we get 
\[
W(e)=W(f)+\int_f^e{M_1(\xi)\,M_2(\xi)\,\Phi(\xi)\,d\,\xi}>0.
\]
On the other hand, $M_1^\prime(e) M_2(e)-M_2^\prime(e) M_1(e)=0,$ so we get the contradiction.
Now let us assume that there exists $g\in(f,\,e)$ such that $M_1(g)=0$ and $M_1^\prime(g)>0.$ Then
\[
W(g)=W(f)+\int_f^g{M_1(\xi)\,M_2(\xi)\,\Phi(\xi)\,d\,\xi}>0.
\] 
On the other hand, $W(g)=M_1^\prime(g) M_2(g)-M_2^\prime(g) M_1(g)<0.$ The contradiction obtained ends the proof.

\begin{lem}\label{Sturm_3}
Suppose that the spectral problem (\ref{geneigen}) has three discrete eigenvalues $\mu_0<\mu_1<\mu_2\leq 0,$ and the corresponding eigenfunctions $M_0(z),\,\,M_1(z),\,\,M_2(z)$ are defined on $(a,b). $ We assume in addition that
\begin{itemize}
\item
$\lim\limits_{z \to a+0  } M_i(z)=\lim\limits_{z \to b-0  } M_i(z)=0, \quad i=0,\,1,\,2;$
\item
there exists $c\in(a,\,b)$ such that $M_2(c)=0,$ $M_2^\prime(c)<0,$ and  $M_2(z)$ does not have another points of intersection with the horizontal axis on the segment $(a,\,b).$
\end{itemize}
Then there does not exist the eigenfunction $M_0(z),$ not identically equal to zero, corresponding to the eigenvalue $\mu_0.$
\end{lem}

{\bf Proof.}
Without loss of generality, we can assume that $M_1\,|_{(a,\,b)}>0$ and $M_2|_{(a,\,c)}>0$ (by virtue of lemmae \ref{Sturm_1}, \ref{Sturm_2} $M_1(z)$ does not intersect the horizontal axis on $(a,\,b)$).  Now assume that $M_0(z)$ is not identically zero on $(a,\,b).$ Then, in accordance with the lemma \ref{Sturm_1}, $M_0(z)$ does not intersect the horizontal axis on this segment (we compare $M_0$ with the function $M_2$) and we can assume in addition that $M_0|_{(a,b)}>0.$
On virtue of the above assumptions, the function 
\[
W(z)=\frac{\mu_1-\mu_0}{\kappa}\int_a^z M_0(\xi)\,M_1(\xi) \Phi(\xi)\,d\,\xi
\]  
is growing and non-negative on the segment $(a,\,b),$ hence $W(b)>0.$ But on the other hand, $W(b)=M_0^\prime(b) M_1(b)-M_1^\prime(b) M_0(b)=0,$ so we get a contradiction.  

\begin{crl}
The following assertions are true:
\begin{itemize}
\item
The  eigenvalue problem (\ref{geneigen}) has at most one discrete eigenvalue $\mu<0,$ corresponding to the nonzero eigenfunction $M(z)$.
\item
If such an eigenvalue does exist, then it is simultaneously the discrete eigenvalue of the operator $\mathcal{L}^s,$ corresponding to the eigenfunction
\[
\left\{M(z),\,s\,M(z)\left(\mu-\frac{\nu+2}{\eta_s(z)^{\nu+3}}\right)^{-1}\right\}^{tr}.
\]
\item
The operator $\mathbb{L}=J\,\cdot\,\mathcal{L}^s$ has at most one discrete eigenvalue lying in $\mathbb{C}^{+}$. 
\item
If such an eigenvalue does exist, then it belongs to $\mathbb{R}^{+}.$
\end{itemize}
\end{crl}

\subsection{The Evans function and spectral stability}

\subsubsection{Introductory remarks}

We are going to formulate the conditions excluding the existence of the discrete eigenvalues of the operator $\mathbb{L}$ belonging to $\mathbb{C}^{+}.$ For this purpose, we use a technique based on some properties of the Evans function \cite{Evans3,Evans4,KaPromis}, an analytic function of the spectral parameter $\lambda\in\mathbb{C}^{+},$ that nullifies on those values of the parameter $\lambda$ which belong to the set $\sigma_{pt}(\mathbb{L}) \cap \mathbb{C}^+.$ Usually, $E(\lambda)$ is defined as a Wronskian constructed on the solutions of a dynamical system  equivalent to the corresponding spectral problem. The  Evans function  most often is studied numerically, but some of its asymptotic properties (essentially used in this paper) can be analyzed analytically.

To begin with, let us remind that, on virtue of lemma \ref{eigenzero}, $\,0\,\in\,\sigma_{pt}( \mathbb{L} ).$ Next, we observe, that the variational equation
\begin{equation}\label{vareq1}
\delta \left(H+s\,Q   \right) \,|_{U_s}=0,
\end{equation}
where $U_s=(w_s(z),\,\eta_s(z))^{tr}$, is equivalent to the traveling wave ODEs  (\ref{varsol1})-(\ref{varsol2}). Differentiating  (\ref{vareq1}) with respect to $s,$ we get:
\[
\mathcal{L}^s\,\partial\,U_s/\partial\,s=-\left(\begin{array}{lc} 0 & 1  \\ 1 & 0 \end{array}\right) U_s.
\]
Multiplying both sides of this equality by $J$ from the left, we obtain:
\[
\mathbb{L}\left( \partial\,U_s/\partial\,s \right)\,=\,-{U_s}^\prime.
\]   
From this we conclude that span$\left\{{U_s}^\prime,\,\partial\,U_s/\partial\,s \right\}$\,$\subset$\,gker($\mathbb{L}$), which, in turn, implies the equality $E^\prime(0)=0.$ If we were able to estimate the $sign\,E^{\prime\prime}(0)$, and also sign\,$E(+\infty)$  (the latter can be done by several standard methods), then from the equality 
\begin{equation}\label{sigmin}
sign{E(+\infty)} \,\cdot\,sign{E^{\prime\prime}(0)}=+1,
\end{equation}
it would follow that the number of intersections of the graph of the function $E(\lambda), \,\,\,\lambda\in\,\mathbb{R}^+$ with the horizontal axis $Re(\lambda)$ should be even. However, since this contradicts the results obtained above (see the corollary at the end of the previous subsection), then there would be no intersections in this case at all. On the contrary, the negativity of the product appearing in the formula (\ref{sigmin}) indicates the existence of an unstable mode.

\subsubsection{The multi-symplectic representation} 

In the evaluation of the sign of $E^{\prime\prime}(0)$ we follow the papers \cite{BriDerks_97,BriDerks_02}. From there, most of the designations are borrowed. The main formula is based on the theory of multi-symplectic systems. We will not fully cover this rather cumbersome theory here, but only concentrate on those fragments that are necessary for deriving the basic formula. Thus, first of all, the Hamiltonian system must be written in the equivalent multi-symplectic form
\begin{equation}\label{multisymgen}
\hat M\,Z_t+\hat K\,Z_x=\nabla S(Z),
\end{equation}
where $Z\in R^{2n}$, $\hat M,\,\hat K$ are $2\,n\,\times\,2\,n$ skew-symmetric constant matrices, $S(Z)$ is a smooth function and $\nabla
$ is the gradient in $R^{2n}.$ The matrices $\hat M,\,\hat K$ generate in $R^{2n}\,\times\,\,R^{2n}$ two-forms
\[
\omega(\zeta_1,\,\zeta_2)=\left(\hat M \zeta_1,\,\zeta_2    \right), \quad 
k(\zeta_1,\,\zeta_2)=\left(\hat K \zeta_1,\,\zeta_2    \right)
\]  
and
\[
\Omega(\zeta_1,\,\zeta_2)=\left(\hat J_s \zeta_1,\,\zeta_2    \right),
\]
where $\hat{J}_s=\hat{K}-s\,\hat{M}.$ It is assumed that $\det \hat{J}_s \neq 0$  and hence the form $\Omega$ is not degenerate. In the multi-symplectic approach the function  $\tilde Z(z; a,\,b,\,s)$ is considered, describing the shape of a multiparameter family of solitary waves and satisfying the dynamical system
\[
\hat J_s \tilde Z^\prime=\nabla V(\tilde Z),
\]
where $V(\cdot)$ is $S(\cdot)$ plus additional features arising from symmetry (we will give the precise definition of them when addressing the system (\ref{MGM1})-(\ref{MGM2})). The linearization $U(z)$ about the solitary wave solutions satisfies the dynamical system
\[
U^\prime(z)=A(z,\,\lambda,a,\,b,\,s) U(z), \quad U\in C^{2\,n}, 
\]  
where $\lambda\in \,\mathbb{C}$ is the spectral parameter,
\[
A(z,\,\lambda,a,\,b,\,s)=\hat{J}_s^{-1}\left\{D^2 V\left( \tilde Z (z;\,a,\,b,\,s \right)-\lambda\,\hat M   \right\}.
\]
It can be shown that the shape function $\tilde Z(z;\,a,\,b,\,s)$ satisfies the variational equation
\[
\frac{\delta}{\delta\,\tilde Z} \left(H(\tilde Z)-s\,I(\tilde Z)   \right)=0,  
\]
where
\[
H(\tilde Z)=\frac{1}{2} \int_{-\infty}^{+\infty}\left[k(\tilde Z,\,\tilde Z^\prime)+2\,V(\tilde Z)   \right]d\,z
\]
is the Hamiltonian function, while
\begin{equation}\label{GenIimpuls}
I(\tilde Z)=\frac{1}{2} \int_{-\infty}^{+\infty} \omega(\tilde Z,\,\tilde Z^\prime)\,d\,z
\end{equation}
is the generalized momentum. In this notation the sign of $E^{\prime\prime}(0)$ is expressed as follows:
\begin{equation}\label{sgnEgen}
sgn\, E^{\prime\prime}(0)=\zeta_{00}^{-} \left[\frac{d\,I}{d\,s}-B(s)\right],
\end{equation}
where $\zeta_{00}^{-}$ and $B(s)$ are expressed in terms of the combination of the vectors $Z_0^{-}(\,a,\,b,\,s)=\lim\limits_{z \to -\infty }\tilde Z(z;\,a,\,b,\,s)$ and $Z_0^{+}(\,a,\,b,\,s)=\lim\limits_{z \to +\infty }\tilde Z(z;\,a,\,b,\,s).$
The multi-symplectic formalism is described below with reference to the system under study.

 \subsubsection{Multi-symplectic representation of the system (\ref{MGM1})-(\ref{MGM2}) and evaluation of the $sign\,E^{\prime\prime}(0)$}

In order to take advantage of the formalism proposed in 
\cite{BriDerks_97,BriDerks_02}, we should write down the initial system in the multi-symplectic form. Introducing new functions
\begin{equation*}
q=\eta^{-(\nu+2)}, \quad \Phi_x=q^{-\frac{1}{\nu+2}}, \quad v=w_x, \quad r_x=w-C_0, \quad  
p=-\Phi_t+\gamma\,w-\kappa\,v_x,
\end{equation*}
we can rewrite (\ref{MGM1})-(\ref{MGM2}) as the first-order system
\begin{equation}\label{multis1}
-\Phi_t-\kappa v_x=p-\gamma\,w,
\end{equation}
\begin{equation}\label{multis2}
\Phi_x=q^{\frac{-1}{\nu+2}},
\end{equation}
\begin{equation}\label{multis3}
w_x=v,
\end{equation}
\begin{equation}\label{multis4}
w_t+q_x=0,
\end{equation}
\begin{equation}\label{multis5}
p_x=0,
\end{equation}
\begin{equation}\label{multis6}
r_x=w-C_0.
\end{equation}
The multi-symplectic form of the system (\ref{multis1})-(\ref{multis6}) is then as follows
\begin{equation}\label{multisym}
\hat M\,Z_t+\hat K\,Z_x=\nabla S,
\end{equation}
where  $Z=\left(w,\,q,\,v,\,\Phi,\,r,\,p\right)^{tr},$
\[
\hat M=\left( \begin{array}{cccccc} 
0 & 0 & 0 & -1 & 0 & 0 \\
0 & 0 & 0 & 0 & 0 & 0  \\
0 & 0 & 0 & 0 & 0 & 0  \\
1 & 0 & 0 & 0 & 0 & 0  \\
0 & 0 & 0 & 0 & 0 & 0  \\
0 & 0 & 0 & 0 & 0 & 0  \\
 \end{array} \right), \quad 
\hat K=\left( \begin{array}{cccccc} 
0 & 0 & -\kappa & 0 & 0 & 0 \\
0 & 0 & 0 & -1 & 0 & 0      \\
\kappa & 0 & 0 & 0 & 0 & 0  \\
0 & 1 & 0 & 0 & 0 & 0       \\
0 & 0 & 0 & 0 & 0 & -1       \\
0 & 0 & 0 & 0 & 1 & 0       \\
 \end{array} \right),
\]
\[
S=p(w-C_0)-\frac{\gamma}{2}w^2-\frac{1}{\alpha}\,q^\alpha+\frac{\kappa}{2} v^2, \quad \alpha=\frac{\nu+1}{\nu+2}.
\]

The next step will be the use of symmetry properties for the purpose of 
constructing {\it a manifold at infinity } $\mathcal{M}(a,\,b)$, \cite{BriDerks_97,BriDerks_02}. The system (\ref{multisym}) is evidently invariant with respect to the translation group $Z \rightarrow Z+ \epsilon (0,\,0,\,0,1,\,0,\,0)^{tr}$, $\epsilon\in \mathbb{R}$ having the generator $\hat X=\partial/\partial \Phi.$ To this symmetry corresponds a pair of functions \cite{BriDerks_97,BriDerks_02} $P=-w,$ $Q=-q$ with the properties
\[
\hat M\,\hat X (Z)=\nabla P(Z), \qquad \hat K\,\hat X (Z)=\nabla Q(Z). 
\] 
In addition, the symmetry of the initial system will be used, which allows one to extend the homoclinic solution to a two-parameter family of analogous solutions. A direct verification shows that the following assertion holds

\begin{lem}
The system (\ref{MGM1})-(\ref{MGM2}) is invariant with respect to the family of transformations:
\begin{equation}\label{extend_sol}
\bar t=e^\mu\,t, \quad \bar x=x, \quad \bar w=e^{-\frac{\nu+1}{\nu+3}\mu}w+A,  \quad \bar \eta=e^{\frac{2}{\nu+3}\mu}\eta,  
\end{equation}
where $\mu,\,A\,\in \mathbb{R}$ are arbitrary parameters.
\end{lem}
 
The above symmetry induces the following group of invariance of the system 
(\ref{VarTW1})-(\ref{VarTW2}) describing the TW solutions:
\begin{equation}\label{TWeq_sym}
\tilde w(z)=e^{-\frac{\nu+1}{\nu+3}\mu}w_s(z)+A, \quad \tilde \eta(z)=e^{\frac{2}{\nu+3}\mu}\eta_s(z), \,\, \tilde \eta_\infty=e^{\frac{2}{\nu+3}\mu}\eta_\infty, \,\,\tilde z=z,\,\, \tilde s=e^{-\mu}\,s.
\end{equation}
Combining the translational symmetry of (\ref{multisym}) with the symmetry of the system (\ref{VarTW1})-(\ref{VarTW2})  which makes it possible to extend the set of homoclinic solutions to a multiparametric family, we can eventually construct a non-degenerate manifold $\mathcal{M}(a,\,b)$, which is necessary for analyzing formula (\ref{sgnEgen}) in our particular case. The vector-function $\tilde Z=\left(\tilde w,\,\tilde q,\, \tilde v,\, \tilde \Phi,\, \tilde r,\, \tilde p    \right)^{tr}$ satisfies the following variational equation  (cf with \cite{BriDerks_02}):
\begin{equation}\label{extendvar}
\left(\hat K-\tilde s\,\hat M\right)\,\tilde Z^\prime=
\nabla S(\tilde Z)-a\nabla P(\tilde Z)-b\nabla Q(\tilde Z),
\end{equation}
which, when written out componentwise, looks as follows
\begin{equation}\label{TWmulti1}
\tilde s\,\tilde \Phi_z-\kappa \tilde v_z=\tilde p+a-\gamma\,\tilde w,
\end{equation}
\begin{equation}\label{TWmulti2}
-\tilde \Phi_z=b-\tilde q^{\frac{-1}{\nu+2}},
\end{equation}
\begin{equation}\label{TWmulti3}
\kappa\,\tilde w_z=\kappa\, \tilde v,
\end{equation}
\begin{equation}\label{TWmulti4}
-\tilde s\, \tilde w_z+\tilde q_z=0,
\end{equation}
\begin{equation}\label{TWmulti5}
-\tilde p_z=0,
\end{equation}
\begin{equation}\label{TWmulti6}
\tilde r_z=\tilde w-C_0.
\end{equation}
Integrating Eq. (\ref{TWmulti2}) over the segment $(-\infty,\,z)$ and using the requirement that the function $\tilde \Phi(z)$ should be bounded, we obtain the condition 
\[
e^\mu=\left(\frac{b}{\eta_\infty}   \right)^{\frac{\nu+3}{2}}.
\]
Integrating (\ref{TWmulti2}) and taking into account the above formula, we get
\begin{equation}\label{t_Phi}
\tilde\Phi(z)=\frac{b}{\eta_\infty}\int_{-\infty}^z \left[\eta_s(\xi)-\eta_\infty   \right]\,d\,\xi+C_1.
\end{equation}
The requirement of the boundedness of the function $\tilde r(z)$ leads to the condition $C_0=A.$ Taking them into account, we get the expression 
\begin{equation}\label{t_r}
\tilde r(z)=\left(\frac{\eta_\infty}{b}\right)^{\frac{\nu+1}{2}}\int_{-\infty}^z w_s(\xi)\,d\,\xi+C_2.
\end{equation}
The remaining functions are expressed as follows:
\begin{equation}\label{t_wqp}
\left\{ \begin{array}{l} \tilde w=\left(\frac{\eta_\infty}{b}\right)^{\frac{\nu+1}{2}}
w_{ s}(z)+a(1+\gamma^{-1}),  \\
\tilde q= \left(\frac{b\,\eta_s(z)}{\eta_\infty}   \right)^{-(\nu+2)},  \\
\tilde v=\left(\frac{\eta_\infty}{b}   \right)^{\frac{\nu+1}{2}} w_s^\prime(z), \\
\tilde p=\gamma\,a, 
\end{array}
\right.
\end{equation}
and, thus, the vector-valued functions $\tilde Z,$ $\tilde Z^\prime$ are represented in the form
\begin{equation}\label{tld_Z}
\begin{split}
\tilde Z= 
\Bigl( \left(\frac{\eta_\infty}{b}\right)^{\frac{\nu+1}{2}}
w_{ s}(z)+a(1+\gamma^{-1}),\,\left(\frac{b\,\eta_s(z)}{\eta_\infty}   \right)^{-(\nu+2)},\,\left(\frac{\eta_\infty}{b}   \right)^{\frac{\nu+1}{2}} w_s(z)^\prime,\,
\\
\qquad \qquad\qquad \qquad \theta(z)+C_1,\,\varphi(z)+C_2,\,\gamma \,a   \Bigr)^{tr}, 
\end{split}
\end{equation}
\begin{equation}\label{tld_Zpr}
\begin{split}
\tilde Z^\prime=
\Bigl(\left(\frac{\eta_\infty}{b}\right)^{\frac{\nu+1}{2}}
w_{ s}(z)^\prime,\,-(\nu+2)  \frac{\eta_\infty^{\nu+2} \eta_s(z)^\prime}{b^{\nu+2}\eta_s^{\nu+3}},\,\left(\frac{\eta_\infty}{b}   \right)^{\frac{\nu+1}{2}} w_s(z)^{\prime\prime}, 
\\
\qquad \qquad\qquad \qquad\frac{b}{\eta_\infty} \left[\eta_s(z)-\eta_\infty   \right],\,\left(\frac{\eta_\infty}{b}\right)^{\frac{\nu+1}{2}} w_s(z),\,0   \Bigr)^{tr}, 
\end{split}
\end{equation}

where
\begin{equation}\label{theta_varphi}
\theta(z)=\frac{b}{\eta_\infty}\int_{-\infty}^{z} \left[\eta(\xi)-\eta_\infty   \right]\,d\,\xi, \quad
\varphi(z)=\left(\frac{b}{\eta_\infty}\right)^{-\frac{\nu+1}{2}}\int_{-\infty}^{z} w_s(\xi)\,d\,\xi. 
\end{equation}
From the formula (\ref{tld_Z}) we calculate the vectors $Z^{\pm}_0$, which are as follows:
\begin{equation}\label{Z_0_minus}
Z_0^{-}=\lim\limits_{z \rightarrow -\infty} \tilde Z(z)=
\left(a (1+\gamma^{-1} ),\,b^{-(\nu+2)},\,0,\,C_1,\,C_2,\,\gamma\,a \right)^{tr},
\end{equation}
\begin{equation}\label{Z_0_plus}
Z_0^{+}=\lim\limits_{z \rightarrow +\infty} \tilde Z(z)=
\left(a (1+\gamma^{-1} ),\,b^{-(\nu+2)},\,0,\,\theta_\infty+C_1,\,\varphi_\infty+C_2,\,\gamma\,a \right)^{tr},
\end{equation}
where $\theta_\infty=\lim\limits_{z \rightarrow +\infty} \theta(z),$ 
$\varphi_\infty=\lim\limits_{z \rightarrow +\infty} \varphi(z).$

Now that almost all the necessary tools have been laid out, we can proceed to an analysis of the sign of the second derivative of the Evans function, which, according to \cite{BriDerks_02}, is expressed by the relation
\begin{equation}\label{secorderE}
E^{\prime\prime}(0)=\chi_{00}^{-} \left( \frac{d}{d\,s}\,I(\tilde Z) -\omega(Z_0^{+},\,\partial_s Z_0^{+}) \right).
\end{equation}
The coefficient $\chi_{00}^{-}$ is obtained from the condition for normalizing the eigenvectors of the matrix $A_\infty=\lim\limits_{z \rightarrow \infty}\,A(z; a,\,b,\,s)$ and the eigenvectors of the adjoining matrix $A_\infty^*$. The computations of this quantity are rather cumbersome, so they are moved to Appendix A, in which the following formula is derived:
\begin{equation}\label{chi_00}
\chi_{00}^{-}=\frac{\left(s\,\eta_\infty^{\nu+3}\right)^2}{2\,\tilde C^2\,\mathcal{D}^{3/2}\,\kappa\,(\nu+2)^2}, \quad \mathcal{D}= \frac{\beta-s^2\,\eta_\infty^{\nu+3}}{\kappa\,(\nu+2)}>0, 
\end{equation}
$\tilde C$ is a constant.
%
%
%
%
%

Thus, we proceed to calculate the remaining terms appearing in formula 
(\ref{secorderE}). Since we are not interested in the whole extended family 
(\ref{extend_sol}), but only in the special case of solutions of system 
(\ref{VarTW1})-(\ref{VarTW2}) satisfying the asymptotic conditions (\ref{asympt}), we carry out calculations for $a=0$ and $b=\eta_\infty.$
The generalized impulse is calculated on the basis of formula (\ref{GenIimpuls}):
\begin{equation}\label{Perl_GenImpuls}
I(\tilde Z)|_{b=\eta_\infty}=\int_{-\infty}^{+\infty} w_s(z)\left[\eta_s(z)-\eta_\infty   \right]d\,z.
\end{equation}
In order to calculate the derivative of the functional $I(\tilde Z)$ with respect to the variable $s$, we need to obtain the explicit dependence of $w_s(z)$ and $\eta_s(z)$ on the velocity. This can be done if we exclude the speed from the system 
(\ref{varsol1})-(\ref{varsol2})  using the scaling $w_s(z)=s^\alpha w_0(z),$ $\eta_s(z)=s^\delta \eta_0(z).$ Indeed, if we put $\alpha=(\nu+1)/(\nu+3),$  $\delta=-2/(\nu+3),$ then we obtain the system
\begin{equation}\label{varsol1_s_less}
  \,w_0-\eta_0^{-(\nu+2)}+\eta_{0,\,\infty}^{-(\nu+2)}   =0,
\end{equation}
\begin{equation}\label{varsol2_s_less}
\left(\gamma-\kappa\,\partial_{z}^2   \right) w_0+\eta_0-\eta_{0,\,\infty} =0,
\end{equation}
which does not contain the parameter $s.$ Thus we have:
\begin{equation}\label{derImpulse}
\frac{d}{d\,s} I(\tilde Z)|_{b=\eta_\infty}=s^{-4/(\nu+3)}\,\frac{\nu-1}{\nu+3}\, 
\int_{-\infty}^{+\infty} w_0(z)\left[\eta_0(z)-\eta_{0,\,\infty}   \right]d\,z. 
\end{equation}
Since the homoclinic loop representing the solitary wave solution lies to the right from the saddle point $(\eta_\infty,\,0)$ then $\eta_s(z)-\eta_{\infty}>0,$ and this implies the inequality  $\eta_0(z)-\eta_{0,\,\infty}>0.$ The inequality $w_0(z)<0$,  in turn, appears directly from Eq.~(\ref{varsol1_s_less}). So the integral in the  formula (\ref{derImpulse}) is negative and the whole expression is positive if $\nu\in\,(-3,\,1).$ 

We must also calculate the expression $\omega(Z_0^{+},\,\partial_s Z_0^{+})$, which appears in formula (\ref{secorderE}). Taking the derivative of (\ref{Z_0_plus}) with respect to $s$, we get:
\[
\partial_s\,Z_0^{+}=\left(0,\,0,\,0,\,-\frac{2}{\nu+3}s^{-\frac{\nu+5}{\nu+3}}\,\int_{-\infty}^{+\infty} \left[\eta_0(z)-\eta_{0,\,\infty}   \right]d\,z, \right. 
\]
\[
\left. \frac{\nu+1}{\nu+3}\,s^{-\frac{2}{\nu+3}}\,\int_{-\infty}^{+\infty} w_0(z)  \,d\,z,\, 0 \right).
\]
Thus,
\[
\omega(Z_0^{+},\,\partial_s Z_0^{+})=-\frac{2}{\nu+3}a\,\left(1+\gamma^{-1} \right)\,
s^{-\frac{\nu+5}{\nu+3}} \int_{-\infty}^{+\infty} \left[\eta_0(z)-\eta_{0,\,\infty}   \right]d\,z,
\]  
which is zero when $a=0.$ Combining (\ref{derImpulse}) with (\ref{hclexist0}) and taking into account that the homoclinic loop exists when $\nu>-1,$ we can formulate the following assertion:
\begin{thm}\label{Mainth}
The solitary wave solution of the system (\ref{MGM1})-(\ref{MGM2}) moving with velocity $s>0$ and having the asymptotics $\lim\limits_{|x|\rightarrow +\infty}w(t,\,x)=0,$ $\lim\limits_{|x|\rightarrow +\infty}\eta(t,\,x)=s^{-\frac{2}{\nu+3}}\,\eta_{0,\,\infty}\,>0$ is  spectrally stable if  $\nu\in(-1,\,1)$, and the inequality (\ref{hclexist0}) holds. 
\end{thm}

Thus, we've established that under the above conditions the operator of the spectral problem (\ref{eigenvalprobl}) does not have eigenvalues belonging to $\mathbb{C}^{+}.$ In connection with this, the question arises: can the result obtained be used to formulate the conditions for the stability of soliton-like TW solutions of the system (\ref{Perl_1A})-(\ref{Perl_1B})? As it was mentioned before, the stability of soliton-like solutions supported by this system was investigated in \cite{VMSS} using the numerical methods which do not give the possibility to obtain any qualitative result. In order to formulate the spectral problem, let us consider the  system (\ref{Perl_1A})-(\ref{Perl_1B}) written in TW coordinates $t,\,z=x-s\,t$: 
\begin{eqnarray}
u_t=s\,u_z-\partial_z\,\left(\gamma-\kappa\partial^2_z\right) \rho^{\nu+2},  \label{Perl_1TWcoord}\\
\rho_t=s\,\rho_z-\rho^2\,u_z, \label{Perl_2TWcoord}
\end{eqnarray}
where we assume, as before, that $\gamma=\beta/(\nu+2)>0,$ $\kappa=-\sigma/(\nu+2)> 0$. In accordance with \cite{VMSS}, we denote the soliton solutions of the system (4) by the symbols $u=u_s(z),\,\,\rho=R_s(z).$
Inserting the perturbations of the form
\[
u(t,\,z)=u_s(z)+\epsilon\,e^{\lambda\,t}\,\hat U(z), \quad 
\rho(t,\,z)=R_s(z)+\epsilon\,e^{\lambda\,t}\,\hat \rho(z)
\]
into (\ref{Perl_1TWcoord})-(\ref{Perl_2TWcoord}) and dropping out the terms of the order $O(\epsilon^2)$, we get the following spectral problem:
\begin{equation}\label{Perl_spectr}
\left\{   \begin{array}{l} \lambda\,\hat U=\partial_z\left[ s\,\hat U-\left(\gamma-\kappa\partial_z^2\right)\,(\nu+2) R_s^{\nu+1}\,\hat\rho\right], 
   \\ \\
 \lambda\,\hat \rho=s\,\partial_z \hat\rho-2\,R_s \,u_s^\prime \hat\rho-R_s^2\partial_z \hat U.    \end{array} \right.  
\end{equation}
 Below we'll show that the following assertion holds:

\begin{thm}
The point spectra of the problems (\ref{Perl_spectr}) and (\ref{eigenvalprobl}) are identical.   
\end{thm}

{\bf Proof.}
In analysing the connection between the two spectral problems, we will use the following easily verifiable identities:
\begin{equation}\label{identities} \begin{array}{l} 
R_s(z)=\eta_s^{-1}(z), \qquad\qquad u_s(z)=\left(\gamma-\kappa\partial^2_z\right)\,w_s(z), \\ \\
\hat \rho(z)=-\frac{1}{\eta_s^{2}(z)}\,N(z), \qquad\qquad \hat U(z)=\left(\gamma-\kappa\partial^2_z\right) \,M(z). \end{array}
\end{equation}
Taking into account the invertibility of the operator $\left(\gamma-\kappa\partial^2_z\right) $ and using the relations (\ref{identities}), we can rewrite  the first equation of the system (\ref{eigenvalprobl}) in the form 
\[
\lambda \left(\gamma-\kappa\partial^2_z\right)^{-1} \hat U=
\left(\gamma-\kappa\partial^2_z\right)^{-1}\,\partial_z \left[s\,U-\left(\gamma-\kappa\partial^2_z\right) \,(\nu+2) R_s^{\nu+1} \hat\rho   \right], 
\]
which is identical with the first equation of the system (\ref{Perl_spectr}). The second equation of the system (\ref{eigenvalprobl}) can be converted in the following way:
\[
\lambda N=\partial_z\left[\left(\gamma-\kappa\partial^2_z\right) M+s\,N   \right]
\]
implies
\[
-\lambda \eta_s^2\,\hat\rho=\partial_z\left[\hat U- s\,\eta_s^2\,\hat\rho   \right],
\]
or
\[
\lambda \hat\rho={R_s^2}\partial_z\left[s\,\frac{1}{R_s^2}\,\hat\rho-\hat U   \right],
\]
or
\[
\lambda \hat\rho={R_s^2}\left[s\left(\frac{1}{R_s^2}\,\partial_z\hat\rho -\frac{2}{R_s^3}R_s^\prime\hat\rho\right)-\partial_z\hat U   \right]= 
 s\,\partial_z\hat\rho-R_s^2\partial_z\hat U-2\,s\,\frac{R_s^\prime}{R_s}\hat\rho,
\]
which is identical with the second equation of the system (\ref{Perl_spectr}) on virtue of the identity $s\,R'_s=R^2_s\,u'_s.$

Moving in the opposite direction, we can rewrite  the equation
\[ \lambda\,\hat U=\partial_z\left[ s\,\hat U-\left(\gamma-\kappa\partial_z^2\right)\,(\nu+2) R_s^{\nu+1}\,\hat\rho\right],
\]
as
\[
\lambda\,\left(\gamma-\kappa\partial^2_z\right)\,M=
\partial_z\left(\gamma-\kappa\partial^2_z\right)\left[ s\,M+\frac{\nu+2}{\eta_s^{\nu+3}}\,N  \right],
\]
which is equivalent to the first equation of the system (\ref{eigenvalprobl}). 

The equation
\[
\lambda\,\hat \rho=s\,\partial_z \hat\rho-2\,R_s \,u_s^\prime \hat\rho-R_s^2\partial_z \hat U\] 
is equivalent to
\[
-\lambda \frac{N}{\eta_s^2}=-s\,\partial_z \,\left( \frac{N}{\eta_s^2}\right)+2\,R_s\,u_s^\prime \left( \frac{N}{\eta_s^2}  \right)- R_s^2\partial_z \left(\gamma-\kappa\partial^2_z\right)\,M,
\]
or
\[
\lambda \frac{N}{\eta_s^2}=s\,\partial_z \,\left( \frac{N}{\eta_s^2}\right)-2\,R_s\,u_s^\prime \left( \frac{N}{\eta_s^2}  \right)+\frac{1}{\eta_s^2}\partial_z \left(\gamma-\kappa\partial^2_z\right)\,M,
\]
or
\[
\lambda \,{N}={\eta_s^2}\left\{s\,\left[\frac{1}{\eta_s^2}\partial_z N-2\,\frac{N}{\eta_s^3}\partial_z\eta_s  \right]+2\,s\,N\frac{\eta_s^\prime}{\eta_s^3} +\frac{1}{\eta_s^2}\left(\gamma-\kappa\partial^2_z\right)\,M \right\},
\]
which is equivalent to the second equation of the system (\ref{eigenvalprobl}).
To obtain the last equality, we took advantage of the identity $u_s^\prime=-s\,\eta^\prime_s.$ 

Since earlier in \cite{VMSS} it was shown that the essential spectrum of the operator appearing in formula (\ref{Perl_spectr}) coincides with the imaginary axis, then on the basis of the results obtained above it is possible to formulate

\begin{crl}
Under the assumptions of the theorem \ref{Mainth}, the soliton-like TW solutions of the system (\ref{Perl_1A})-(\ref{Perl_1B}) are spectrally stable.
\end{crl}

\section{Discussion}

Thus, we have shown that fulfillment of the inequality $\nu\,\in\,(-1,\,1)$ assures that the soliton-like TW solutions to the system (\ref{Perl_1A})-(\ref{Perl_1B}) describing the wave of rarefaction are spectally stable. The result obtained in this work is in agreement with that obtained numerically in paper \cite{VMSS}. In conclusion, we would like to note that the presence of higher derivatives in the first equation of the system (\ref{Perl_1A})-(\ref{Perl_1B}) is related to the spatial nonlocality in the lowest approximation. In this connection it is of interest to consider the problem in the following approximation and to trace how the inclusion of additional terms affects the dynamics and stability of soliton-like solutions.

\section*{ Appendix A}

In order to trace the behavior of the vectors $\tilde Z,\,\,\tilde
Z^\prime$ for large values of the arguments, we consider the linearization of the dynamical system 
\begin{equation}\label{App_DS1}
\left\{\begin{array}{l} \frac{d}{d\,z}\eta_s=\dot \eta_s, \\  \\
\frac{d}{d\,z}\dot\eta_s=\frac{\eta_s^{\nu+3}}{\kappa(\nu+2)}\left[
\kappa(\nu+2)(\nu+3)\dot\eta_s^2/\eta_s^{\nu+4}-s^2(\eta_s-\eta_\infty)
+\gamma \frac{\eta_s^{\nu+2}-\eta_\infty^{\nu+2}}{\left(\eta_s\,\eta_\infty\right)^{\nu+2}}\right], \end{array} \right.
\end{equation}
which is equivalent to (\ref{HDS}). The linear part of the system (\ref{App_DS1}) in variables $x=\eta_s-\eta_\infty,\,\,\,y=\dot\eta_s$ will have the following form:
\begin{equation}\label{App_DS_lin}
\left(\begin{array}{c} x  \\ y \end{array}\right)^\prime=
\left(\begin{array}{cc} 0 & 1  \\ \mathcal{D} & 0 \end{array}\right)\,\left(\begin{array}{c} x  \\ y \end{array}\right),
\end{equation}
where $\mathcal{D}=
\frac{1}{\kappa (\nu+2)}\,\left[\beta-s^2\,\eta_\infty^{\nu+3}\right]>0.$ Thus, for $|z|\gg 1$ we obtain the asymptotics 
\[
x=\eta_s-\eta_\infty \cong\left\{\begin{array}{c}  \tilde C e^{\sqrt{\mathcal{D}}\,z}, \qquad z\ll -1, \\ \\ \tilde C e^{-\sqrt{\mathcal{D}}\,z}, \qquad z \gg 1. \end{array} \right.
\]
Constants at the exponential functions are the same because of the symmetry of the homoclinic trajectory with respect to the horizontal axis. Using the first equation of the system (\ref{varsol1}), we get the asymptotics for another component of the homoclinic solution:
\[
w_s \cong -\frac{\nu+2}{s\,\eta_\infty^{\nu+3}}\left\{\begin{array}{c}  \tilde C e^{\sqrt{\mathcal{D}}\,z}, \qquad z \ll-1,  \\  \\ \tilde C e^{-\sqrt{\mathcal{D}}\,z}, \qquad z \gg 1. \end{array} \right.
\]   
From this we get the asymptotics (cf with \cite{BriDerks_02}):
\[
\begin{split}
\Psi^{\pm}=\lim\limits_{z \rightarrow \pm \infty}e^{\pm \sqrt{\mathcal{D}} z} \tilde Z^\prime=\tilde C \sqrt{\mathcal{D}}
&\left[  \pm\left(\frac{\eta_\infty}{b}\right)^{\frac{\nu+1}{2}} \frac{\nu+2}{s\,\eta_\infty^{\eta+3}};\, \pm  \left(\frac{\eta_\infty}{b}\right)^{\nu+2} \frac{\nu+2}{\eta_\infty^{\eta+3}};\,
-\sqrt{\mathcal{D}}\left(\frac{\eta_\infty}{b}\right)^{\frac{\nu+1}{2}} \frac{\nu+2}{s\,\eta_\infty^{\eta+3}}; \right. \\
&\left. \frac{b}{\eta_\infty\,\sqrt{\mathcal{D}}};\,
-\left(\frac{\eta_\infty}{b}\right)^2 \frac{\nu+2}{s\,\eta_\infty^{\eta+3}\sqrt{\mathcal{D}}};\,0 
    \right]. 
\end{split}
\]
The coefficient $\chi_{00}^{-}$ is obtained from the normalization condition 
\[
1=\left(\hat J_s\,\eta_1^{-},\,\Psi^{+}\right),
\]
where $\eta_1^{-}=\chi_{00}^{-}\Psi^{-},$ $\hat J_s=\hat K-s\,\hat M$ (see \cite{BriDerks_02}, section 3).
In the general case, the above formula is very cumbersome, but in the case  we are interested in,  i.e., when $b=\eta_\infty$, a more straightforward expression emerges:
\[
1=\chi_{00}^{-}\,\left(\hat J_s \Psi^{-}  \right)^{tr}\,\Psi^{+}=
\]
\[
=\tilde C^2\,\mathcal{D}\,\chi_{00}^{-}\left[  
\frac{\kappa (\nu+2)\mathcal{D}+s^2\eta_\infty^{\nu+3}}{s\eta_\infty^{\nu+3}\sqrt{\mathcal{D}}};-\frac{1}{\sqrt{\mathcal{D}}};-\frac{\kappa(\nu+2)}{s\eta_\infty^{\nu+3}};0;0;-\frac{\nu+2}{s\sqrt{\mathcal{D}}\eta_\infty^{\nu+3}} \right]\,*
\]
\[
*\left[\frac{\nu+2}{s\eta_\infty^{\nu+3}};\frac{\nu+2}{\eta_\infty^{\nu+3}};
-\frac{\sqrt{\mathcal{D}}(\nu+2)}{s\eta_\infty^{\nu+3}};\frac{1}{\sqrt{\mathcal{D}}};
-\frac{\nu+2}{s\eta_\infty^{\nu+3}\sqrt{\mathcal{D}}};0\right]^{tr}=
\frac{2\chi_{00}^{-}\tilde C^2\,\mathcal{D}^{3/2}\,\kappa\,(\nu+2)^2}{\left(s\,\eta_\infty^{\nu+3}\right)^2}.
\]
Hence
\[
\chi_{00}^{-}=\frac{\left(s\,\eta_\infty^{\nu+3}\right)^2}{2\,\tilde C^2\,\mathcal{D}^{3/2}\,(\nu+2)^2\kappa}>0.
\]

\section*{ Appendix B}

Here we analyze the fulfillment of the hypotheses from \cite{BriDerks_02}, which guarantee the existence of the normalization of the Evans function, under which $\lim\limits_{\lambda \rightarrow +\infty}E(\lambda)=1.$ Substituting into  Eq. (\ref{multisym}) a perturbation of the form 
$Z(t,\,z)=Z(z)+\varepsilon\,e^{\lambda\,t}\,U(z),$  performing elementary algebraic transformations, and dropping the higher-order terms  in $\varepsilon$, we get the linear  dynamical system:
\begin{equation}\label{PertMult}
U^\prime=\hat A(z;\,\lambda)\,U(z), \qquad U\in R^{2\,n},
\end{equation}
where 
\[
\hat A(z;\,\lambda)=\hat J_s^{-1}\left(\hat B(z)-\lambda\,\hat M   \right)=\left( \begin{array}{cccccc} 
0 & 0 & 1 & 0 & 0 & 0 \\
-\lambda & 0 & s & 0 & 0 & 0  \\
\gamma/\kappa & -s\,R(z)/\kappa & 0 & -\lambda/\kappa & 0 & -1/\kappa  \\
0 & -R(z) & 0 & 0 & 0 & 0  \\
1 & 0 & 0 & 0 & 0 & 0  \\
0 & 0 & 0 & 0 & 0 & 0  \\
 \end{array} \right),
\]
$\hat B(z)=D^2 S(Z),$ $R(z)=\eta_s^{\nu+3}(z)/(\nu+2).$ The matrix  $\hat A(z;\,\lambda)$ is related to a pair of constant matrices 
\[\hat A^{\pm}(\lambda)=\lim\limits_{z\rightarrow \pm\infty} \hat A(z;\,\lambda).\] In our case  
\[
A^{+}(\lambda)=A^{-}(\lambda)=A_{\infty}(\lambda)=
\left( \begin{array}{cccccc} 
0 & 0 & 1 & 0 & 0 & 0 \\
-\lambda & 0 & s & 0 & 0 & 0  \\
\gamma/\kappa & -s\,R_\infty/\kappa & 0 & -\lambda/\kappa & 0 & -1/\kappa  \\
0 & -R_\infty & 0 & 0 & 0 & 0  \\
1 & 0 & 0 & 0 & 0 & 0  \\
0 & 0 & 0 & 0 & 0 & 0  \\
 \end{array} \right),
\]
where $R_\infty=\eta_\infty^{\nu+3}/(\nu+2).$ The spectral problem for the matrix $A_{\infty}(\lambda)$ can be written as follows:
\begin{equation}\label{eigAinf}
det\left[A_{\infty}(\lambda)-\mu\,I  \right]=\mu^2 \left[\kappa\mu^4-\gamma\mu^2+R_\infty (\lambda-s\mu)^2   \right]=0.
\end{equation}
We prove the following assertions  necessary for applying the results of \cite{BriDerks_02} to the investigation of the asymptotics of the Evans function at infinity.

\begin{lem} The spectrum of the matrix $A_{\infty}(0)$ contains a pair of real eigenvalues $\pm \zeta \neq 0;$ whereas the remaining eigenvalues nullify.

If $\lambda\in \mathbb{C}^{+},$ then the spectrum of $A_{\infty}(\lambda)$ contains a pair of eigenvalues $\mu_1^-,\,\mu_2^-$ with negative real parts, while the remaining  eigenvalues have non-negative real parts. 
\end{lem}

{\bf Proof.}
The first assertion is obvious, since for $\lambda=0$ it is not difficult to calculate the eigenvalues from the formula (\ref{eigAinf}):
\[
\mu_{1,\,2}=\pm\sqrt{\frac{\beta-s^2\,\eta_\infty^{\nu+3}}{\kappa (\nu+2)}}=\pm \zeta, 
\] 
while $\mu_{3,...6}=0.$
To prove the second assertion, we apply the method of asymptotic expansions, representing the eigenvalues in the form  of a series $\mu=a_0+a_1\,\lambda+....$ Substituting this expression in (\ref{eigAinf}) and equating the coefficients of the corresponding powers of $\lambda$ to zero, we obtain a  system of algebraic equations. In view of the awkwardness of the computations, we used the {\it Mathematica} package for deriving these equations. Thus, nullifying  zero-order coefficients, we get the equation  
\begin{equation}\label{Mpacksys0}
a_0^2\left[a_0^2\kappa+\left(s^2\,R_\infty-\gamma\right)\right]=0,
\end{equation}
having the pair of nonzero solutions
\[
a_{0}^\pm=\pm\sqrt{\frac{\beta-s^2\,\eta_\infty^{\nu+3}}{\kappa (\nu+2)}}=\pm \zeta.
\]

Equating to zero the coefficient of $\lambda^1$, we get:
\begin{equation}\label{Mpacksys1}
2\,a_0\,\left[a_1\left(s^2\,R_\infty-\gamma  \right)+2\,a_0^2 \,a_1\kappa-s\,R_\infty\right]=0.
\end{equation}
For $a_0\neq 0$ Eq. (\ref{Mpacksys1}) gives the expression
\[
a_1\,|_{a_0^{\pm}}=\frac{s\eta_\infty^{\nu+3}}{\beta-s^2\eta_\infty^{\nu+3}}.
\]
So for $0<\lambda<<1,$ we have the following pair of roots:
\[
\mu_{1}^-=-\zeta+a_1\,\lambda +O(\lambda^2)<0
\]
and
\[
\mu_{1}^+=\zeta+a_1\,\lambda +O(\lambda^2)>0.
\]

The second pair of roots, corresponding to $a_0=0,$ is obtained from the following approximation. Putting the coefficient 
of $\lambda^2$ to be equal to zero, we obtain the equation
\begin{equation}\label{Mpacksys3}
R_\infty (s a_1-1)^2-a_1^2\,\gamma=0,
\end{equation}
whose solutions are expressed as follows:
\[
a_{11}=\frac{s\,R_\infty-\sqrt{\gamma R_\infty}}{s^2 R_\infty-\gamma}>0,
\]
\[
a_{12}=\frac{s\,R_\infty+\sqrt{\gamma R_\infty}}{s^2 R_\infty-\gamma}<0.
\]
Using these solutions, we obtain the second pair of the roots:
\[
\mu_{2}^{-}=a_{12}\,\lambda+O(\lambda^2)<0,
\]
\[
\mu_{2}^{+}=a_{11}\,\lambda+O(\lambda^2)>0.
\]
Since the characteristic Eq.~(\ref{eigAinf}) always has a pair of zero solutions, the above construction exhausts all possible cases corresponding to small values of the parameter $\lambda>0.$ And this is enough to complete the proof of the second point because of the fact that, as can easily be seen,  $Re(\mu_i)$ change signs only when the parameter $\lambda$ belongs to the imaginary axis. Thus, Eq.~(\ref{eigAinf}) will have exactly two solutions with negative real part for any $\lambda$ with positive real part. 


\section*{ Acknowledgements} The authors gratefully acknowledge   Maxim Pavlov for paying attention at the transformation leading to the Hamiltonian representation of the source system. We are also greatly indebted  to  Sergij Kuzhel for valuable discussions during the preparation of this manuscript. One of the authors (VV) acknowledges support from the Polish Ministry of Science and Higher
Education.

\end{document}